\documentclass[superscriptaddress,preprint]{revtex4}%
\usepackage{amsfonts}
\usepackage{amsmath}
\usepackage{amssymb}
\usepackage{graphicx}%
\begin{document}
\title{Pseudo-magnetoexcitons in strained graphene bilayers without external magnetic fields}
\author{Zhigang Wang}
\affiliation{LCP, Institute of Applied Physics and Computational Mathematics, P. O. Box
8009, Beijing 100088, China}
\author{Zhen-Guo Fu}
\affiliation{SKLSM, Institute of Semiconductors, CAS, P. O. Box 912, Beijing 100083, China}
\affiliation{LCP, Institute of Applied Physics and Computational Mathematics, P. O. Box
8009, Beijing 100088, China}
\author{Fawei Zheng}
\affiliation{LCP, Institute of Applied Physics and Computational Mathematics, P. O. Box
8009, Beijing 100088, China}
\author{Ping Zhang}
\affiliation{LCP, Institute of Applied Physics and Computational Mathematics, P. O. Box
8009, Beijing 100088, China}
\affiliation{Beijing Computational Science Research Center, Beijing 100089, China}

\begin{abstract}
\textbf{The structural and electronic properties of graphene leads its charge
carriers to behave like relativistic particles, which is described by a
Dirac-like Hamiltonian. Since graphene is a monolayer of carbon atoms, the
strain due to elastic deformations will give rise to so-called `pseudomagnetic
fields (PMF)' in graphene sheet, and that has been realized experimentally in
strained graphene sample. Here we propose a realistic strained graphene
bilayer (SGB) device to detect the pseudo-magnetoexcitons (PME) in the absence
of external magnetic field. The carriers in each graphene layer suffer
different strong PMFs due to strain engineering, which give rise to Landau
quantization. The pseudo-Landau levels (PLLs) of electron-hole pair under
inhomogeneous PMFs in SGB are analytically obtained in the absence of Coulomb
interactions. Based on the general analytical optical absorption selection
rule for PME, we show that the optical absorption spectrums can interpret the
corresponding formation of Dirac-type PME. We also predict that in the
presence of inhomogeneous PMFs, the superfluidity-normal phase transition
temperature of PME is greater than that under homogeneous PMFs.}

\end{abstract}
\maketitle
\author{Zhigang Wang}
\affiliation{LCP, Institute of Applied Physics and Computational Mathematics, P. O. Box
8009, Beijing 100088, China}
\author{Zhen-Guo Fu}
\affiliation{SKLSM, Institute of Semiconductors, CAS, P. O. Box 912, Beijing 100083, China}
\affiliation{LCP, Institute of Applied Physics and Computational Mathematics, P. O. Box
8009, Beijing 100088, China}
\author{Fawei Zheng}
\affiliation{LCP, Institute of Applied Physics and Computational Mathematics, P. O. Box
8009, Beijing 100088, China}
\author{Ping Zhang}
\thanks{To whom correspondence should be addressed. Email address: zhang\_ping@iapcm.ac.cn}
\affiliation{LCP, Institute of Applied Physics and Computational Mathematics, P. O. Box
8009, Beijing 100088, China}


Graphene has been an optimal playground to realize the exciting ideas in
condensed matter physics \cite{Novoselov,Castro, Goerbig} since the carriers
in graphene behave like massless Dirac particles. Many efforts have been
devoted to detecting the striking electronic properties of graphene, such as
Klein paradox \cite{Katsnelson}, anomalous quantum Hall effect \cite{ZhangY},
and so on. However, it was soon realized that the structural and mechanical
properties should also be important both from theoretical interest and from
applications of graphene \cite{Geim}, such as the strain engineering in
graphene
\cite{Pereira,Levy,Vozmediano,Guinea1,Guinea,Bahat-Treidel,Juan,Verberck}. It
has been shown recently that specific forms of strain produce a strong PMF in
graphene, which effectively break the time-reversal symmetry \cite{Vozmediano}%
. The strain-induced PMF is expected to produce PLLs, and consequently, the
quantum Hall effect, even in the absence of external magnetic field
\cite{Guinea1}. These intriguing properties extend to graphene multilayers.
Recently, the shear mode in graphene multilayers has been observed in Ramman
spectroscopy experiment \cite{Tan}, and the high temperature Bose-Einstein
condensation and superfluidity of indirect excitons or electron-hole pairs in
Graphene $n$-$p$ bilayers \cite{Iyengar,Lozovik1,Berman,Berman1} are also
predicted. Differing from other bilayer $n$-$p$ systems, such as coupled
semiconductor quantum wells \cite{Lozovik,Snoke,Eisenstein}, $n$-$p$ type SGB
separated by a dielectric layer should be an ideal setup to create PME due to
the PMF induced by strain rather than the applied magnetic filed. Because of
the difference of ripples or elastic deformations in each graphene layer
sample on a substrate, in fact, it is more realistic to fabricate different or
say inhomogeneous PMFs in electron and hole layers, which is hard to realize
with external magnetic fields. \begin{figure}[ptb]
\begin{center}
\includegraphics[width=.6\linewidth]{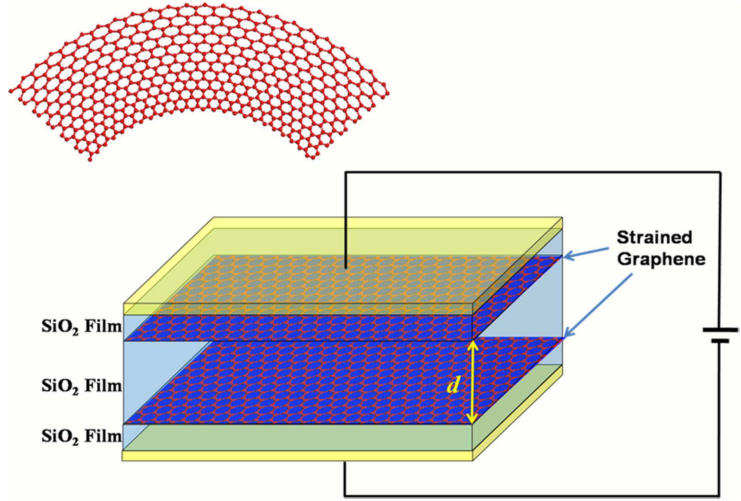}
\end{center}
\caption{(Color online) \textbf{Basic Scheme of the proposed SGB device.} Two
strained graphene monolayers are separated by dielectric spacer. Electron/hole
carriers in each layer are created by $n(p)$-type doping or applied external
gates. Indirect PME can be generated by PMFs in SBG. Uniform PMF perpendicular
to the graphene surface is created in a circular arc strained graphene
ribbon.}%
\end{figure}

On the basis of these interesting and reasonable ideas, in the present work,
we will theoretically analyze the PME properties in SGB device without
applying magnetic field. In the following, we show that the formation of PME
in SGB could be determined from the optical absorption selection rule of PME,
which is related to the imbalance parameter $\gamma\mathtt{=}\sqrt{B_{s}%
^{h}/B_{s}^{e}}$ of the strain-induced PMFs $B_{s}^{h}$ and $B_{s}^{e}$
suffered by Dirac holes and electrons in SGB. We also find that comparing with
homogeneous case ($\gamma$=$1$), the Kosterlitz-Thouless (KT) transition
critical temperature will be improved by a factor of $\left[  \left(
2\mathtt{+}\gamma\mathtt{+}\gamma^{-1}\right)  /4\right]  ^{2}$. Moreover, we
suggest some technical skills for designing the SGB setup that can make use of
this realization to detect PME.

As illustrated in Fig. 1, the system we considered here is that of two
parallel strained graphene layers separated by an insulating slab of SiO$_{2}%
$. By varying the chemical potential via tuning the bias voltages of the two
gates located near the corresponding graphene sheets, the spatially separated
electrons and holes in different layers can be generated. In order to obtain
the analytical PLL expression, we suppose both graphene sheets are bent to
circular arc \cite{Guinea}, which can lead to PMFs $B_{s}^{e/h}=8\beta
c^{e/h}/a_{0}$ (in unit of $\hbar/e\mathtt{\equiv}1$), and the corresponding
fictitious gauge field vector potentials can be chosen as $\boldsymbol{A}%
_{s}^{e/h}\mathtt{=}B_{s}^{e/h}\left(  0,x^{e/h}\right)  $. Here, $a_{0}$ is
the lattice constant, $\beta$=$-\partial\ln t/\partial\ln a_{0}\mathtt{\approx
}2$ and $t$ denotes the nearest-neighbor hopping parameter,\ and $c^{e/h}$ is
the numerical constant representing the strength of strain in electron/hole
layers. In general $c^{e}\mathtt{\neq}c^{h}$, and thus $B_{s}^{e}\mathtt{\neq
}B_{s}^{h}$ in SGB device, which results in the fact that $\gamma$ departs
from $1$.

This SGB system is described by the Hamiltonian $H$=$H_{0}$+$U\left(
\boldsymbol{r}^{e}\mathtt{-}\boldsymbol{r}^{h}\right)  $, where%
\begin{equation}
H_{0}\mathtt{=}\hbar v_{f}\left[  \pi_{x}^{e}\sigma_{1}\mathtt{\otimes}%
\sigma_{0}\mathtt{-}\pi_{y}^{e}\sigma_{2}\mathtt{\otimes}\sigma_{0}%
\mathtt{+}\pi_{x}^{h}\sigma_{0}\mathtt{\otimes}\sigma_{1}\mathtt{-}\pi_{y}%
^{h}\sigma_{0}\mathtt{\otimes}\sigma_{2}\right]  , \label{H1}%
\end{equation}
with Fermi velocity $v_{f}\mathtt{\approx}10^{6}$ m$\mathtt{\cdot}$s$^{-1}$,
$\boldsymbol{\pi}^{e/h}\mathtt{=}\boldsymbol{p}^{e/h}\mathtt{\mp}\frac{e}%
{c}\boldsymbol{A}_{s}^{e/h}$, and Pauli matrices $\sigma_{i}$ ($i\mathtt{=}%
0,1,2$). $U\left(  \boldsymbol{r}^{e}\mathtt{-}\boldsymbol{r}^{h}\right)
$=$-e^{2}/\epsilon\sqrt{|\boldsymbol{r}^{e}\mathtt{-}\boldsymbol{r}^{h}%
|^{2}\mathtt{+}d^{2}}$ is Coulomb interaction (CI) between the pair of
spatially separated electron and hole, where $d$ is the spacer thickness and
$\epsilon$ ($\sim4.5$ for SiO$_{2}$) denotes the dielectric constant of
spacer. In the following text, the relative CI strength notation is taken as
$\lambda\mathtt{\equiv}\left(  e^{2}/\epsilon\right)  /(\hslash v_{f})$.
Taking some coordinate transformations, the free part of Hamiltonian $H$ can
be written as
\begin{equation}
H_{0}\mathtt{=}\frac{\hbar v_{f}\left(  l_{e}+l_{h}\right)  }{\sqrt{2}\left(
l_{e}l_{h}\right)  ^{3/2}}\left(
\begin{array}
[c]{cccc}%
0 & l_{h}c_{+} & l_{e}c_{-}^{\dagger} & 0\\
l_{h}c_{+}^{\dag} & 0 & 0 & l_{e}c_{-}^{\dagger}\\
l_{e}c_{-} & 0 & 0 & l_{h}c_{+}\\
0 & l_{e}c_{-} & l_{h}c_{+}^{\dag} & 0
\end{array}
\right)  , \label{H2}%
\end{equation}
with $l_{e/h}\mathtt{=}\sqrt{\hbar/eB_{s}^{e/h}}$ the pseudomagnetic lengths
in electron/hole layer (in following text, we set the length unit as
$\sqrt{l_{e}l_{h}}$) and the harmonic lowering operators $c_{\pm}$ (see
Supplementary Information part I).\begin{figure}[ptb]
\begin{center}
\includegraphics[width=.6\linewidth]{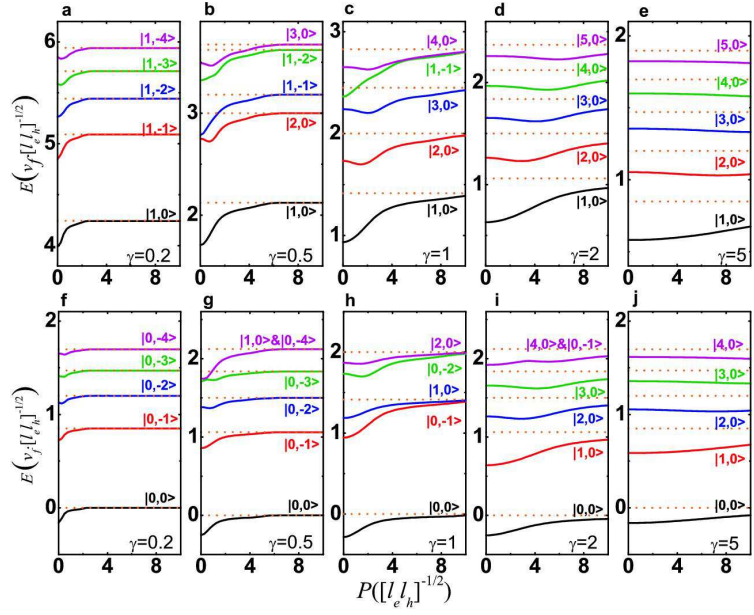}
\end{center}
\caption{(Color online) \textbf{Dispersions of the first few PLLs for fully
occupied case (upper panels) and partially occupied (lower panels) case, as
functions of $P$ with different $\gamma$.} The dashed horizontal lines denote
the non-interacting PLLs. The parameters are chosen as $\mu$=$0$, $\lambda
$=$0.49$, $d$=$0.2$, and $R$=$0$.}%
\end{figure}

To obtain eigenvalues $E_{n_{+},n_{-}}$ of the total Hamiltonian $H$, the CI
part $U$ needs to be treated as a perturbation in the first order $_{0}\langle
n_{+}^{\prime},n_{-}^{\prime}|U|n_{+},n_{-}\rangle_{0}$, where $|n_{+}%
,n_{-}\rangle_{0}$ is the eigenstates of $H_{0}$ corresponding to the
eigenvalues, i.e., the PLLs of $H_{0}$ without CI
\begin{equation}
E_{n_{+},n_{-}}^{(0)}\text{=}\frac{1}{\sqrt{2}}\left[  s_{+}\sqrt{|n_{+}%
|}\left(  1\mathtt{+}\gamma^{-1}\right)  \mathtt{-}s_{-}\sqrt{|n_{-}%
|}(1\mathtt{+}\gamma)\right]
\end{equation}
in units of $\hslash v_{f}/\sqrt{l_{e}l_{h}}$, with $s_{\pm}\mathtt{=}%
$sgn$\left(  n_{\pm}\right)  $. The imbalance parameter $\gamma$ will reduce
the degeneracy of PLLs of $H_{0}$. For instance, $|1,0\rangle_{0}$ and
$|0,-1\rangle_{0}$ are degenerate when $\gamma$=$1$, while these two states
turn to be nondegenerate when $\gamma\mathtt{\neq}1$. The PLLs $E_{n_{+}%
,n_{-}}$ of PMEs are not only dependent on $\gamma$, but also related to the
effective momentum $\boldsymbol{P}\mathtt{=}2\frac{l_{e}\boldsymbol{p}%
^{e}\mathtt{+}l_{h}\boldsymbol{p}^{h}}{l_{e}\mathtt{+}l_{h}}$ and coordinate
$\boldsymbol{R}\mathtt{=}\frac{l_{h}\boldsymbol{r}^{e}\mathtt{+}%
l_{e}\boldsymbol{r}^{h}}{l_{e}\mathtt{+}l_{h}}$ of the mass center of PME
since $U$ is related to $\boldsymbol{P}$ and $\boldsymbol{R}$ after
transformations (see Supplementary Information part I). The location of the
chemical potential will determine the attainable possible PLL indices for
electrons and holes. For convenience, in this work we choose the notation
$\mu$ describing the highest filled PLL and consider two cases: (i) the
electron-PLLs with $n_{+}$%
$>$%
$\mu$ are unoccupied and the hole-PLLs with $n_{-}\mathtt{\leqslant}\mu$ are
fully occupied; and (ii) the electron-PLLs with $n_{+}\mathtt{>}\mu$ are
unoccupied and the hole-PLLs with $n_{-}\mathtt{<}\mu$ are fully occupied,
while the PLL at $\mu$ is partially occupied.

The spectrums of the lowest five PLLs as a function of $\boldsymbol{P}$ are
demonstrated in Fig. 2. Figures 2a$\mathtt{-}$2e correspond to the fully
occupied cases while Figs. 2f$\mathtt{-}$2j are for the partially occupied
cases. The PME dispersion strongly depends on the imbalance parameter $\gamma$
(see Fig. 2). On one hand, if the difference in the elastic deformations
between electron and hole layers is large, i.e., $\gamma$ departs from
$\gamma$=$1$ greatly, the PME dispersion will accelerate the transition
process from density wave forms to particle-hole pair forms with increasing
the momentum $\boldsymbol{P}$. On the other hand, when $\gamma$ takes a
moderate value around $1$, the PME dispersion behaves like that of
magnetoexcitons in perfect graphene bilayers under an external magnetic field.
Besides, we also found that, taking the partially occupied cases as an
example, at much larger $\gamma$ the lowest PLLs behave like electron type
$|n_{+},0\rangle$ (for example, Fig. 2j) while at much smaller $\gamma$ the
lowest PLLs behave like hole type $|0,n_{-}\rangle$ (for example, Fig. 2f).
Differing from the magnetoexcitons in an external magnetic field, the PME
dispersion is also related on $\boldsymbol{R}$. By increasing $R$, the PME
dispersion at smaller $P$\ turns to become a particle-hole pair form rather
than a density-wave form (Supplementary Figure S1). \begin{figure}[ptb]
\begin{center}
\includegraphics[width=.6\linewidth]{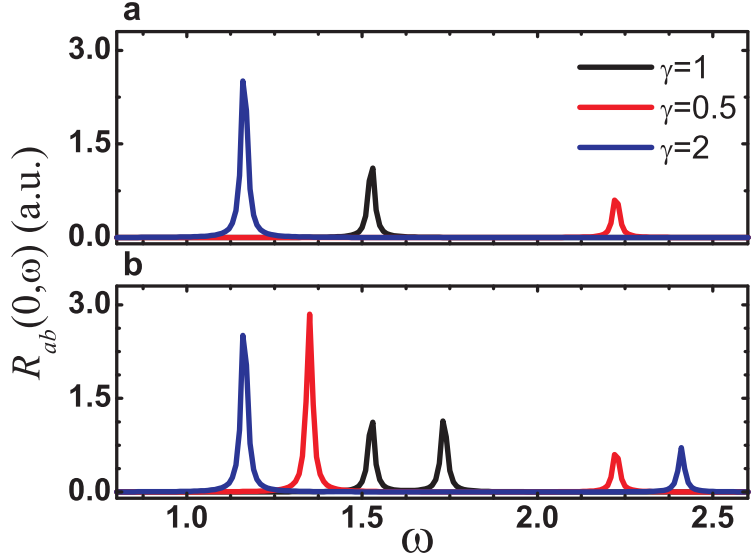}
\end{center}
\caption{(Color online) \textbf{The optical absorption curves of PMEs with
$\mu$=$0$.} \textbf{a}, Fully and \textbf{b} partially occupied cases as
functions of photon energy $\omega$ ($\hbar$=$1$). The spacer thickness is set
to $d$=$0.2$. $\lambda$=$0.49$.}%
\end{figure}


The optical absorption spectroscopy analysis is an instructive method in
studying the properties of magnetoexcitons, because it can determine the
knowledge of magnetoexciton's energies and wave functions at $\boldsymbol{P}%
$=$0$. The particle-hole excitation energy should be revealed directly by the
resonant peaks in the magneto-optical absorption spectroscopy $R_{ab}%
(R,\omega)\mathtt{\varpropto}\sum_{n_{+},n_{-}}\left\vert \langle n_{+}%
,n_{-}|v_{x}|\mu,\mu\rangle\right\vert ^{2}$. In the absence of inter-PLL CI
which is much weaker than intra-PLL CI, for linear polarized light
$\boldsymbol{A}$=$A\mathbf{\hat{x}}$, the optical absorption selection rule
for PMEs is analytically expressed as
\begin{align}
\langle n_{+},n_{-}|v_{x}|\mu,\mu\rangle &  \mathtt{=}c_{1}\delta
_{|n_{+}|,|\mu|}\delta_{|n_{-}|,|\mu|-1}\mathtt{+}c_{2}\delta_{|n_{+}%
|,|\mu|-1}\delta_{|n_{-}|,|\mu|}\nonumber\\
&  \mathtt{+}c_{3}\delta_{|n_{+}|,|\mu|}\delta_{|n_{-}|,|\mu|+1}%
\mathtt{+}c_{4}\delta_{|n_{+}|,|\mu|+1}\delta_{|n_{-}|,|\mu|}, \label{oas}%
\end{align}
where $v_{x}$=$\partial H_{0}/\hbar\partial p_{x}$ is the velocity operator
for free part of Hamiltonian, and the coefficients $c_{1}\mathtt{=-}%
(1\mathtt{+}\gamma)s_{-}(s_{\mu}s_{+}\mathtt{+}1)C$, $c_{2}\mathtt{=}%
(1\mathtt{+}\gamma^{-1})(s_{\mu}\mathtt{+}s_{-})C$, $c_{3}\mathtt{=-}%
(1\mathtt{+}\gamma)(s_{\mu}\mathtt{+}s_{+})C$, and $c_{4}\mathtt{=}%
(1\mathtt{+}\gamma^{-1})s_{+}(s_{\mu}s_{-}\mathtt{+}1)C$ with $C\mathtt{=}%
v_{f}(\sqrt{2})^{\delta_{n_{+},0}+\delta_{n_{-},0}+2\delta_{\mu,0}-6}$ (see
Supplementary Information part II). This optical absorption selection rule
formula (\ref{oas}) offers a better way to analyze PME absorption resonant
peaks in spectroscopy of SGB device. Notice that the optical absorption
formula (\ref{oas}) also depends on $\boldsymbol{R}$ since the PME spectrum is
related to $\boldsymbol{R}$. (For briefness, we just consider the $R$=$0$ case
in following text.)

Especially, for the fully occupied case with $\mu\mathtt{\geqslant}0$, the
selection rule can be further simplified as
\begin{equation}
\langle n_{+},n_{-}|v_{x}|\mu,\mu\rangle=c_{4}\delta_{n_{+},\mu+1}%
\delta_{n_{-},\mu}, \label{13}%
\end{equation}
which indicates that there is only one resonance peak occurring in optical
absorption spectrum, which corresponds to the formation of PME. For example,
the selection rule for fully occupied case of $\mu$=$0$ with the ground state
$|0,0\rangle$ promises that just one resonance peak that corresponds to the
formation of PME state $|1,0\rangle$ occurs, see Fig. 3a. From Fig. 3a one can
also clearly observe that with the imbalance parameter $\gamma$ decreasing
(increasing) from $1$, the resonance peak moves towards high (low) frequency
region, see the red (blue) curve in Fig. 3a.

This effect of inhomogeneity of PMFs can also be seen for the partially
occupied cases, which is illustrated in Fig. 3b. According to the general
selection rule (\ref{oas}), the nonzero elements of $\langle n_{+},n_{-}%
|v_{x}|0,0\rangle$ only occurs at two PMEs states $|1,0\rangle$ and
$|0,-1\rangle$, and thus in this case, there are two resonance peaks occurring
in Fig. 3b. Interestingly, with decreasing (increasing) $\gamma$ from 1, the
resonance peak of $|0,-1\rangle$ moves towards low (high) frequency region, in
contrast, the resonance peak of $|1,0\rangle$ moves towards high (low)
frequency region. The results of cases with $\mu\mathtt{\neq}0$ are similar to
these of $\mu$=$0$, and an example for $\mu$=$1$ is shown in Supplementary
Figure S2.

At this stage, we have not considered the effect of the inter-PLL CIs on the
optical absorption of the PMEs. Because the inter-PLL CI mixes the
noninteracting PLLs, the absorption phenomenon should appear at the energy of
every PME states. Consequently, additional resonance peaks, except of the main
peaks without the inter-PLL CI, will appear in the optical absorption spectrum
(not shown here). The magnitudes of these additional resonant peaks are very
tiny and tend to disappear by increasing the spacer thickness $d$. Therefore,
based on the optical absorption selection rule analyzed above, one can see
that the optical techniques should allow detection of PME locally as well as
the CI effects in the present SGB device. \begin{figure}[ptb]
\begin{center}
\includegraphics[width=.6\linewidth]{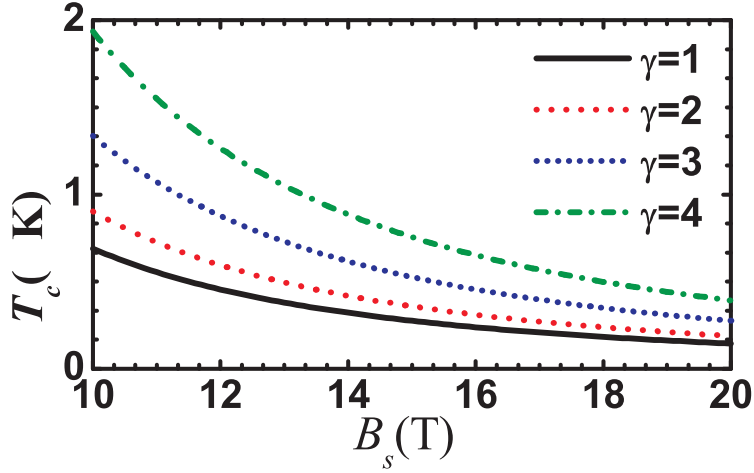}
\end{center}
\caption{(Color online)\textbf{ Calculated KT temperature versus the PMF.} The
density of PME is set as $n$=$4.0\times10^{11}$cm$^{-2}$ and the spacer width
$d$=$30$ nm.}%
\end{figure}

PMEs are also ideal objects in exploring the Bose-Einstein condensation since
they behave as neutral bosons at low densities. Motivated by its importance
both from basic point of interest and from application of graphene-based
electronics, now we turn to address this issue by presenting an attempt at the
theoretical evaluation of effective mass the effective mass $m_{B_{s}}$
($B_{s}\mathtt{\equiv}\sqrt{B_{s}^{e}B_{s}^{h}}$ herein) and superfluid-normal
state, i.e., KT transition temperature $T_{c}$ of PME in this SGB system. As
an example, let us now consider the case of PME on the ($1$, $1$)-PLL, with
the PME energy at small magnetic momenta $E_{1,1}(P)\mathtt{=}\varepsilon
_{B_{s}}^{(b)}(d)\mathtt{+}\frac{P^{2}}{2m_{B_{s}}(d)}$ (see Supplementary
Information part III and ref. 26). In the special limit of $d\mathtt{\gg}%
\sqrt{l_{e}l_{h}}$, one can explicitly obtain $\varepsilon_{B_{s}}%
^{(b)}\mathtt{=}\mathtt{-}\frac{e^{2}}{\epsilon d}$, which is same as that of
magnetoexcitons in external magnetic field since $\varepsilon_{B_{s}}%
^{(b)}(d)$ is independent of $\gamma$. Whereas the effective mass of PME
should be%
\begin{equation}
m_{B_{s}}\mathtt{=}\left[  4/\left(  2\mathtt{+}\gamma\mathtt{+}\gamma
^{-1}\right)  \right]  ^{2}\frac{\epsilon}{4c^{2}}d^{3}B_{s}^{2},
\end{equation}
where the prefactor $\left[  4/\left(  2\mathtt{+}\gamma\mathtt{+}\gamma
^{-1}\right)  \right]  ^{2}$ will reduce to unity in the case of homogeneous
PMFs (i.e., $\gamma\mathtt{=}1$), thus that the effective mass of PME in SBG
is $\left[  4/\left(  2\mathtt{+}\gamma\mathtt{+}\gamma^{-1}\right)  \right]
^{2}$ times of that of magnetoexcitons in perfect bilayer graphene under an
external magnetic field. It is this difference between the effective masses
that results in the difference in the KT transition temperature $T_{c}$, which
reads
\begin{align}
T_{c}  &  \text{=}\left[  \left(  \sqrt{\frac{32}{27}\left[  \left(
sm_{B_{s}}k_{B}T_{c}^{0}\right)  /\left(  \pi\hbar^{2}n\right)  \right]
^{3}\text{+}1}\text{+}1\right)  ^{1/3}\right. \\
&  \left.  -\left(  \sqrt{\frac{32}{27}\left[  \left(  sm_{B_{s}}k_{B}%
T_{c}^{0}\right)  /\left(  \pi\hbar^{2}n\right)  \right]  ^{3}\text{+}%
1}\text{\texttt{-}}1\right)  ^{1/3}\right]  \frac{T_{c}^{0}}{2^{1/3}%
},\nonumber
\end{align}
where the auxiliary quantity is $T_{c}^{0}$=$\frac{1}{k_{B}}\left(  \frac
{\pi\hbar^{2}n\mu_{0}^{2}}{6s\zeta(3)m_{B_{s}}}\right)  ^{1/3}$ with $\mu_{0}%
$=$\frac{\pi\hslash^{2}n}{sm_{B_{s}}\log\left[  s\hslash^{4}\epsilon^{2}/(2\pi
nm_{B_{s}}^{2}e^{4}d^{4})\right]  }$ being the chemical potential of system,
$s$=$4$ the spin degeneracy of PMEs in SGB, and $\zeta(3)$ the Riemann zeta
function. Here, $n=n_{s}+n_{n}$ with $n_{n}$ the normal component density, and
$n_{s}$ the superfluid density. Because $\mu_{0}$ approximately inverse of the
effective mass, a rough consequence can be immediately obtained that the
critical temperature of KT transition of PME in SGB will be greater $\left[
\left(  2\mathtt{+}\gamma\mathtt{+}\gamma^{-1}\right)  /4\right]  ^{2}$ times
than that of magnetoexciton in bilayer graphene in an external magnetic field
\cite{Berman}. Typical results are exhibited in Fig. 4. Thereby, one can
conclude that increasing the imbalance of the strains ($\gamma$ departs from 1
sufficiently) in different layers in SGB device may be one effective method to
promote the KT transition temperature of PME.


In order to design the required SGB device in experiment, firstly, one should
deposit graphene ribbons onto a elastic substrate and then deform it by
bending it into a circular arc automatically (Fig. 1). Following that, we can
transfer it onto the SiO$_{2}$ thin film and then throw off the elastic
substrate by heating or dissolving it via chemical agent carefully. A uniform
PMF could be created in graphene ribbons deposited on SiO$_{2}$ thin film
since the strain distribution in the substrate would project onto graphene
ribbon. In reality, the PMFs arisen from strain in upper and lower bent
graphene ribbons, however, may be different because of different automatic
deformations of them. This is why we discuss the importance of the imbalance
parameter $\gamma$ here. After that one can grow or evaporate metal electrode
on the other side of SiO$_{2}$ thin film, and finally package the device for
measurements. The experimental capability to produce high quality graphene
sample and the feasibility to create strong PMF in graphene, together with the
technical advances for designing graphene multilayer devices make it possible
to perform very original optical and superfluidity-normal states KT transition
experiments of PME in graphene without external magnetic field. Moreover, the
SGB device proposed here could also be used to detect the (valley polarized)
unconventional fractional quantum Hall effect of charged Dirac-type
electron-hole fluid or a Bose condensate of PME in the absence of external
magnetic field by measuring transport coefficients and that, since fixed PME
number corresponds to fixed valley polarization in the SGB device, transport
studies of the SGB system will provide unique information which cannot be
observed by direct study of conventional two-component
electron-electron/electron-hole systems \cite{Eisenstein,Laughlin}.

In summary, in this work we propose a realistic SGB device, and based on the
physics analyzed here, this setup can be made to detect the PME which is
associated with the elastic deformations or curvature in graphene. The cases
discussed in this paper are only illustrative examples of the feasible setup
and applications of the SGB device, and the technical skills mentioned above
are achievable in current experimental capabilities. Therefore, we hope these
predictions and the suggested SGB device can be realized sooner rather than later.

\begin{acknowledgments}
This work was supported by Natural Science Foundation of China under Grants
No. 90921003, No. 10904005, and No. 11004013, and by the National Basic
Research Program of China (973 Program) under Grant No. 2009CB929103.
\end{acknowledgments}

\textbf{Author contributions} All authors equally contributed to this work.

\textbf{Competing financial interests} The authors declare no competing
financial interests.

\textbf{Author Information} Correspondence and requests for materials should
be addressed to P. Z. (zhang\_ping@iapcm.ac.cn).

\newpage

\begin{center}
{\textbf{\emph{Supplementary Information} \newline Pseudo-magnetoexcitons in
strained graphene bilayers without external magnetic fields}\newline Zhigang
Wang,$^{1}$ Zhen-Guo Fu,$^{2,1}$ Fawei Zheng,$^{1}$ and Ping Zhang$^{1,3}%
$\newline\emph{$^{1}$LCP, Institute of Applied Physics and Computational
Mathematics, P. O. Box 8009, Beijing 100088, China\newline$^{2}$SKLSM,
Institute of Semiconductors, CAS, P. O. Box 912, Beijing 100083,
China\newline$^{3}$Beijing Computational Science Research Center, Beijing
100089, China}}
\end{center}

\textbf{I.} Let us explain first how to derive the Hamiltonian (2) of the main
text. Taking the coordinate transformations
\begin{align}
\boldsymbol{P}  &  \mathtt{=}2\frac{l_{e}\boldsymbol{p}^{e}\mathtt{+}%
l_{h}\boldsymbol{p}^{h}}{l_{e}\mathtt{+}l_{h}},\boldsymbol{p}\mathtt{=}%
\frac{l_{e}\boldsymbol{p}^{e}\mathtt{-}l_{h}\boldsymbol{p}^{h}}{l_{e}%
\mathtt{+}l_{h}},\tag{S1}\\
\boldsymbol{R}  &  \mathtt{=}\frac{l_{h}\boldsymbol{r}^{e}\mathtt{+}%
l_{e}\boldsymbol{r}^{h}}{l_{e}\mathtt{+}l_{h}},\boldsymbol{r}\mathtt{=}%
2\frac{l_{h}\boldsymbol{r}^{e}\mathtt{-}l_{e}\boldsymbol{r}^{h}}%
{l_{e}\mathtt{+}l_{h}}, \tag{S2}%
\end{align}
we can rewrite $h_{0}^{e/h}$ in Eq. (1) as
\begin{equation}
h_{0}^{e/h}\mathtt{=}\left\{  \kappa_{e/h}\left(  \frac{P_{x}}{2}\mathtt{\pm
}p_{x}\right)  \mathtt{+}i\left[  \left(  \frac{P_{y}}{2}\mathtt{\pm}%
p_{y}\right)  \mathtt{\mp}\xi\left(  X\mathtt{\pm}\frac{x}{2}\right)  \right]
\right\}  , \tag{S3}%
\end{equation}
where $\kappa_{e/h}\mathtt{=}\left(  l_{e}\mathtt{+}l_{h}\right)  /\left(
2l_{e/h}\right)  $, and $\xi\mathtt{=}1/\left(  2l_{e}l_{h}\right)  $. Here,
$\gamma\mathtt{=}l_{e}/l_{h}\mathtt{=}\sqrt{B_{s}^{h}/B_{s}^{e}}$ describes
the inhomogeneity of PMFs suffered by electrons and holes in different layers.
In the limit of $\gamma$=$1$, the problem of PMEs in SGB is reduced to that of
PMEs in a uniform PMF, which are consistent with those on the magnetoexcitons
in an external magnetic field \cite{S1,S2}. And then one can introduce the
transformation $S\mathtt{=}e^{iXy/l_{e}l_{h}}$, for which $S^{\dag}%
P_{x}S\mathtt{=}P_{x}\mathtt{+}y/l_{e}l_{h}$, $S^{\dag}p_{y}S\mathtt{=}%
p_{y}\mathtt{+}X/l_{e}l_{h}$, and shift $\boldsymbol{r}\mathtt{\rightarrow
}\boldsymbol{r}\mathtt{-}l_{e}l_{h}\boldsymbol{\hat{z}}\mathtt{\times
}\boldsymbol{P}$, $h_{0}^{e/h}$ can be further written as
\begin{equation}
h_{0}^{e/h}\mathtt{=}\kappa_{e/h}\left\{  \left(  \mathtt{\pm}p_{x}%
\mathtt{+}\xi y\right)  \mathtt{\pm}i\left(  p_{y}\mathtt{\mp}\xi x\right)
\right\}  . \tag{S4}%
\end{equation}
The relative coordinate of electron and hole in the $x$-$y$ plane
$\boldsymbol{r}^{e}\mathtt{-}\boldsymbol{r}^{h}$ becomes $\boldsymbol{\tilde
{r}}\mathtt{=}\frac{1}{2}\left(  \gamma\mathtt{-}\gamma^{-1}\right)
\boldsymbol{R}\mathtt{+}\frac{1}{4}\left(  2\mathtt{+}\gamma\mathtt{+}%
\gamma^{-1}\right)  \left(  \boldsymbol{r}\mathtt{-}l_{e}l_{h}\mathbf{\hat{z}%
}\mathtt{\times}\boldsymbol{P}\right)  $. Finally we define the harmonic
lowering operators
\begin{equation}
a\text{=}\sqrt{l_{e}l_{h}}p_{x}\mathtt{-}i\frac{x}{2\sqrt{l_{e}l_{h}}%
},b\text{=}\sqrt{l_{e}l_{h}}p_{y}\mathtt{-}i\frac{y}{2\sqrt{l_{e}l_{h}}},
\tag{S5}%
\end{equation}
and their combination
\begin{equation}
c_{\pm}=\pm(a\mathtt{\pm}ib)/\sqrt{2}. \tag{S6}%
\end{equation}
Substituting them into Eq. (1), we can obtain the Hamiltonian (2) of the main
text. The eigenvalues, i.e., the PLLs of $H_{0}$ without CI are given by Eq.
(3) and the corresponding eigenvectors are given by%
\begin{equation}
\left\vert n_{+},n_{-}\right\rangle _{0}\mathtt{=}2^{\eta}\left(
\begin{array}
[c]{c}%
s_{+}s_{-}\Phi_{|n_{+}|-1,|n_{-}|-1}(\boldsymbol{r})\\
s_{-}\Phi_{|n_{+}|,|n_{-}|-1}(\boldsymbol{r})\\
s_{+}\Phi_{|n_{+}|-1,|n_{-}|}(\boldsymbol{r})\\
\Phi_{|n_{+}|,|n_{-}|}(\boldsymbol{r})
\end{array}
\right)  , \tag{S7}%
\end{equation}
where $\eta\mathtt{=}\frac{\delta_{n_{+},0}+\delta_{n_{-},0}-2}{2}$, $s_{\pm
}\mathtt{=}$sgn$\left(  n_{\pm}\right)  $, and $\Phi_{n_{1},n_{2}%
}(\boldsymbol{r})$ $\mathtt{=}\frac{2^{-\frac{|l_{z}|}{2}}n_{\_}%
!e^{-il_{z}\phi}\delta\left(  l_{z}\right)  r^{|l_{z}|}}{\sqrt{2\pi
n_{1}!n_{2}!}}L_{n_{-}}^{|l_{z}|}(\frac{r^{2}}{2})e^{-\frac{r^{2}}{4}}$ with
$l_{z}$=$n_{1}\mathtt{-}n_{2}$, $n_{-}\mathtt{=}\min(n_{1},n_{2})$,
$\delta\left(  l_{z}\right)  $=sgn$(l_{z})^{l_{z}}\mathtt{\rightarrow}1$ for
$l_{z}$=$0$, and $L\left(  x\right)  $ the Laguerre polynomial. To obtain
eigenvalues of the total Hamiltonian $H$, we need to solve the following
equation:%
\begin{align}
0  &  \mathtt{=}\det\left\Vert \delta_{n_{+},n_{+}^{\prime}}\delta
_{n_{-},n_{-}^{\prime}}(E_{n_{+},n_{-}}^{(0)}\mathtt{-}E)\right. \nonumber\\
&  \left.  \mathtt{+}_{0}\langle n_{+}^{\prime},n_{-}^{\prime}|U\left(
\mathbf{\tilde{r}}\right)  \left\vert n_{+},n_{-}\right\rangle _{0}\right\Vert
. \tag{S8}%
\end{align}
Here, the intra-PLL component of the CI is defined as $_{0}\langle n_{+}%
,n_{-}|U\left\vert n_{+},n_{-}\right\rangle _{0}$, while the inter-PLL
component is defined as $_{0}\langle n_{+}^{\prime},n_{-}^{\prime}|U\left\vert
n_{+},n_{-}\right\rangle _{0}$, where $\left\vert n_{+}^{\prime},n_{-}%
^{\prime}\right\rangle _{0}\neq\left\vert n_{+},n_{-}\right\rangle _{0}$. The
PLLs as a function of $R$ are presented in supplementary Figure S1.
\begin{figure}[ptb]
\includegraphics{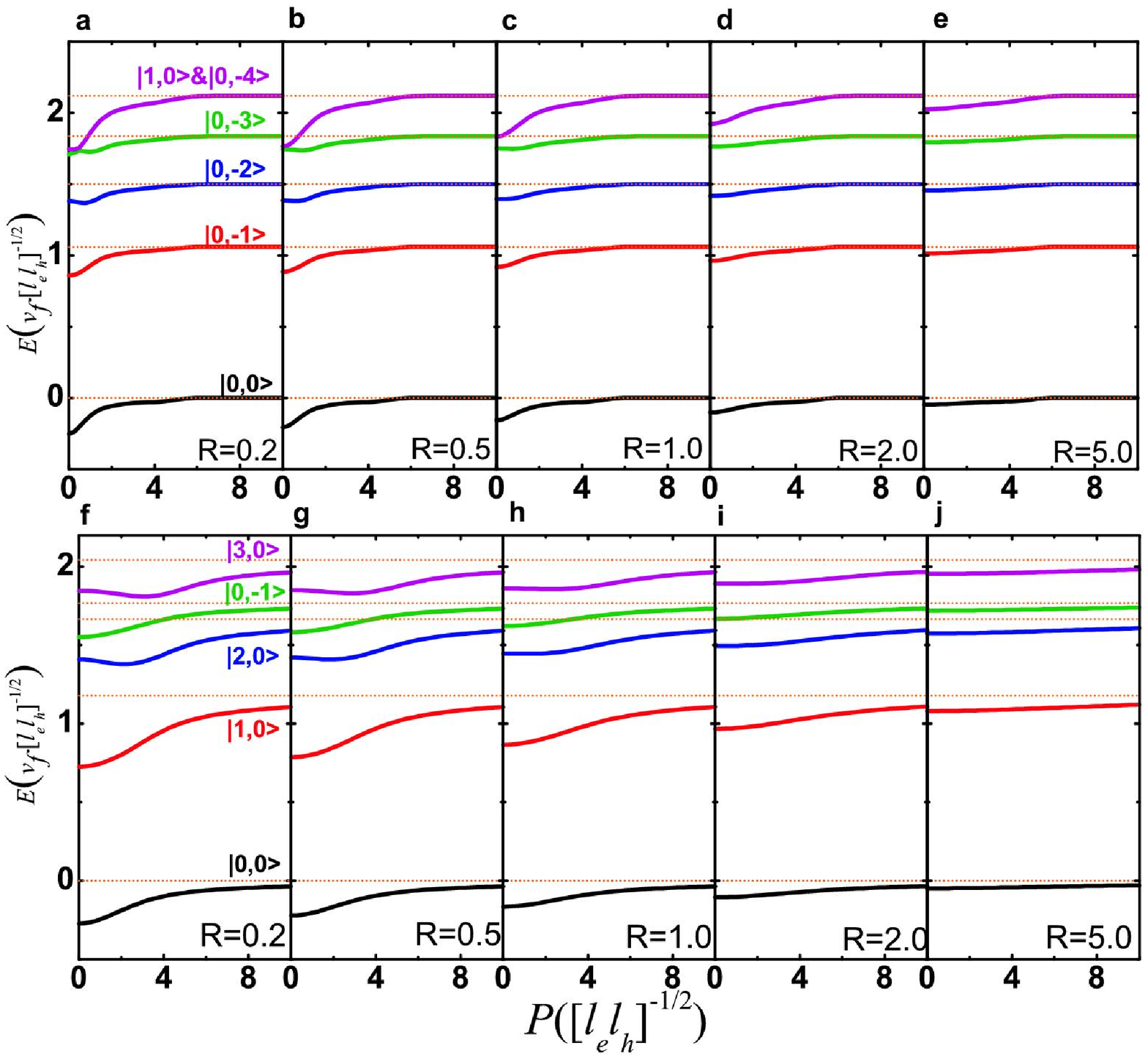}
\par
\begin{center}
\begin{flushleft}
Supplementary Figure S1: \textbf{Dispersions of the first few PLLs as functions
of $P$ with different $R$.} The imbalance parameter $\gamma$=$0.5$ for
\textbf{a}-\textbf{e} while $\gamma$=$1.5$ for \textbf{f}-\textbf{j}, and
other parameters are same as Figure 2 in main text.
\end{flushleft}
\end{center}
\par
\label{S1}\end{figure}

\begin{figure}[ptb]
\includegraphics{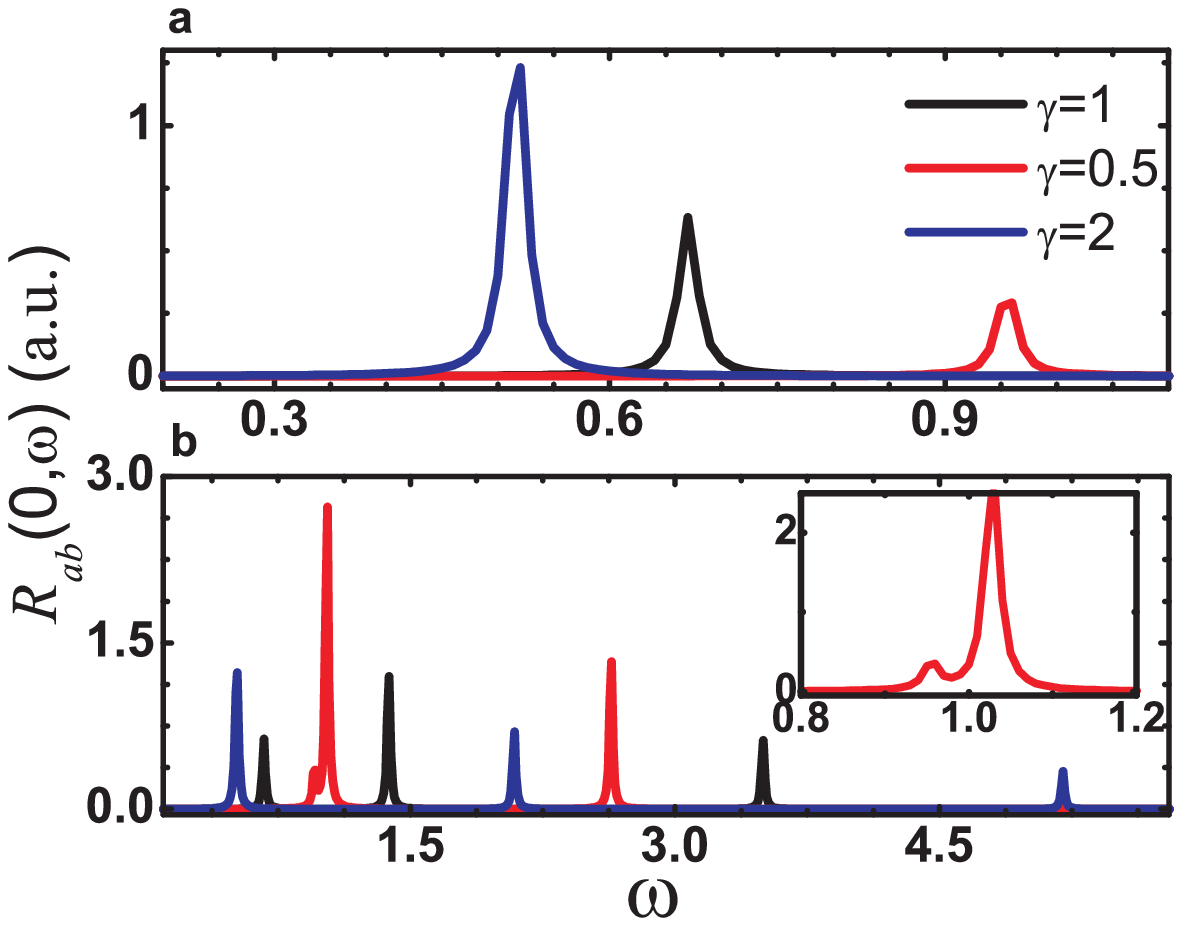} \begin{flushleft}
Supplementary Figure S2: \textbf{The optical absorption curves of PMEs with
$\mu$=$1$. } \textbf{a}, Fully and \textbf{b} partially occupied cases as
functions of photon energy $\omega$ ($\hslash$=$1$). The spacer thickness is
set to $d$=$0.2$. $\beta$=$0.49$. The inset in \textbf{b} is the enlarged
dispersion curves between $\omega$=$0.8$ and $\omega$=$1.2$.
\end{flushleft}
\label{S2}\end{figure}

\textbf{II}. From the Fermi's golden rule, the optical absorption for photons
of frequency $\omega$ yields
\begin{equation}
R_{ab}(R,\omega)\text{=}\frac{2\pi}{\hbar}\sum_{\alpha}\left\vert
\langle\alpha|\frac{e}{c}\boldsymbol{A}\mathtt{\cdot}\boldsymbol{v}%
|0\rangle\right\vert ^{2}\delta(\varepsilon_{eh}^{(\alpha)}(R)\mathtt{-}%
\varepsilon_{0}\mathtt{-}\hslash\omega), \tag{S9}%
\end{equation}
where $\left\vert \alpha\right\rangle $ is the set of quantum numbers
describing a particle-hole excitation and $|0\rangle\mathtt{\equiv}|\mu
,\mu\rangle$ is the ground state in absence of particle-hole excitations with
$\varepsilon_{0}$ being the corresponding ground state energy. $\boldsymbol{v}%
$=$\partial H_{0}/\hslash\partial\boldsymbol{p}$ is the velocity operator for
free part of Hamiltonian, and $\boldsymbol{A}$ is the vector potential. For
linear polarized light $\boldsymbol{A}$=$A\mathbf{\hat{x}}$, choosing a
Lorentzian type broadening, one has%
\begin{equation}
R_{ab}(R,\omega)\mathtt{\varpropto}\sum_{\alpha}\left\vert \langle\alpha
|v_{x}|0\rangle\right\vert ^{2}\frac{\Gamma/2}{(\varepsilon_{eh}^{(\alpha
)}(R)\mathtt{-}\varepsilon_{0}\mathtt{-}\hslash\omega)^{2}\mathtt{+}\Gamma
^{2}/4}, \tag{S10}%
\end{equation}
where $\Gamma$ denotes the linewidth, and $\left\vert \langle\alpha
|v_{x}|0\rangle\right\vert ^{2}\mathtt{=}\sum_{n_{+}>\mu}\sum_{n_{-}%
\leqslant\mu}\int Td\boldsymbol{r}$ is for full occupation while $\sum
_{n_{+}\geqslant\mu}\sum_{n_{-}\leqslant\mu}\int Td\boldsymbol{r}$ is for
partial occupation. The parameter $T$ is defined as%
\begin{equation}
T\mathtt{=}|C_{n_{+},n_{-}}^{\alpha}\langle n_{+},n_{-}|v_{x}|\mu,\mu
\rangle|^{2}, \tag{S11}%
\end{equation}
where $C_{n_{+},n_{-}}^{\alpha}$ is the projection of PME state $\left\vert
\alpha\right\rangle $ on the basis state $|n_{+},n_{-}\rangle$. In the absence
of inter-PLL CI which is much weaker than intra-PLL CI, the PME state
$\left\vert \alpha\right\rangle $ is one special basis state and $C_{n_{_{+}%
},n_{_{-}}}^{\alpha}$=$\delta_{\alpha,(n_{_{+}},n_{_{-}})}$, same as that in
the noninteracting case. As a result, we can analytically obtain the optical
absorption selection rule for PMEs, Eq. (4) in main text.

As an example, we additionally plot the optical absorption spectrum with $\mu
$=$1$ fully and partially occupied cases in supplementary Figure S2. The
selection rule for fully occupied case with $\mu$=$1$ promises that there is
only one resonant peak to occur, which corresponds to the formation of PME
state $|2,1\rangle$ by absorbing a photon quanta $\hbar\omega$. For partially
occupied case, however, there are three non-zero elements according to
selection rule (4), which are $|1,0\rangle$, $|1,-2\rangle$, and $|2,1\rangle
$. Therefore, there are three resonant peaks appearing in optical absorption
spectrum, see supplementary Figure S2b. Similar to those in Fig. 3 in main
text for the $\mu$=$0$ cases, the imbalance of strain-induced PMFs changes the
location of the resonant peak. When the imbalance parameter $\gamma
\mathtt{\neq}1$, these resonant peak moves in a more complex way.

\textbf{III}. In the limit of $d\mathtt{\rightarrow}\infty$ and $B_{s}%
\mathtt{\rightarrow}\infty$, the PME energy can be approximated by only
considering its zeroth order energy part $E_{n,m}^{(0)}$. However, if the PMF
is about $10\mathtt{\sim}20$T, CI can be treated as a perturbation since CI
$e^{2}/\epsilon d$ is just several times less than $\hslash v_{f}/\sqrt
{l_{e}l_{h}}$. In the absence of inter-PLL CI, by substituting the approximate
relation \cite{S3}
\begin{equation}
\langle\langle nmP|U(\boldsymbol{\tilde{r}})|nmP\rangle\rangle\mathtt{=}%
\varepsilon_{nm}^{(b)}\mathtt{+}\frac{P^{2}}{2M_{mn}(B_{s},d)} \tag{S12}%
\end{equation}
for $R\mathtt{=}0$ into Eq. (S8), we can get the asymptotic dispersion law of
a PME for small magnetic momenta. Here the notation $\langle\langle
nmP|U(\mathbf{\tilde{r}})|nmP\rangle\rangle$ denotes the averaging by the
two-dimensional harmonic oscillator eigenfunctions $\Phi_{n,m}(\mathbf{r})$,
and $\varepsilon_{nm}^{(b)}$ is the binding energy, and finally one can obtain
the Eqs. (6) and (7) in main text.

\end{document}